\documentstyle[12pt]{article}

\newcommand\as{\alpha_{\mathrm{S}}}

\def\ep{\epsilon}
\def\ee{$e^+e^-$}
\def\beq{\begin{equation}}
\def\eeq{\end{equation}}
\def\beeq{\begin{eqnarray}}
\def\eeeq{\end{eqnarray}}
\def\cm{{\cal M}}
\def\phase{e^{- i \lambda_{ij} \pi}}
\def\phasot{e^{- i \lambda_{12} \pi}}
\def\phastt{e^{- i \lambda_{23} \pi}}
\def\phasto{e^{- i \lambda_{31} \pi}}

\def\bom#1{{\mbox{\boldmath $#1$}}}
\def\to{\rightarrow}

\def\RS{{\scriptscriptstyle\rm R\!.S\!.}}
\def\cdr{{\scriptscriptstyle\rm C\!.D\!.R\!.}}

\def\CDR{{\rm CDR}}

\newcommand{\la}{\langle}
\newcommand{\ra}{\rangle}

\def\ID{1 \kern -.45 em 1}

\begin{document}

\begin{titlepage}
\renewcommand{\thefootnote}{\fnsymbol{footnote}}
\begin{flushright}
     CERN--TH/98-42 \\ LPTHE--ORSAY 97/57 \\ hep-ph/9802439 
     \end{flushright}
\par \vspace{10mm}
\begin{center}
{\Large \bf
The Singular Behaviour of QCD Amplitudes \\[1ex] 
at Two-loop Order}
\end{center}
\par \vspace{2mm}
\begin{center}
{\bf Stefano Catani}\footnote{On leave of absence from I.N.F.N,
Sezione di Firenze, Florence, Italy.}\\

\vspace{5mm}

{Theory Division, CERN, CH 1211 Geneva 23, Switzerland} \\
and \\
LPTHE, Universit\'{e} Paris-Sud, B\^{a}timent 211, F-91405 Orsay Cedex, France

\vspace{5mm}

\end{center}

\par \vspace{2mm}
\begin{center} {\large \bf Abstract} \end{center}
\begin{quote}
\pretolerance 10000

We discuss the structure of infrared singularities in on-shell QCD
amplitudes at two-loop order. We present a general factorization formula that 
controls all the $\ep$-poles of the dimensionally regularized amplitudes.
The dependence on the regularization
scheme is considered
and the coefficients of the $1/\ep^4, 1/\ep^3$ and $1/\ep^2$ poles are
explicitly given in the most general case. The remaining single-pole
contributions are also explicitly evaluated in the case of amplitudes with a 
$q{\bar q}$ pair.
 
\end{quote}

\vspace*{\fill}
\begin{flushleft}
     CERN--TH/98-42 \\  LPTHE--ORSAY 97/57 \\   February 1998
\end{flushleft}
\end{titlepage}

\renewcommand{\thefootnote}{\fnsymbol{footnote}}

\section{Introduction}
\label{intro}

Perturbative QCD predictions [\ref{book}] are important to test QCD and to
measure its fundamental parameters (e.g. the strong coupling $\as$), to
estimate Standard Model backgrounds for new-physics signals, and to 
understand the interplay between perturbative and non-perturbative phenomena.

Calculations at the leading order (LO) in perturbation theory are  
straightforward: the matrix elements at tree level can be easily computed
by using 
helicity techniques and colour-subamplitude decompositions [\ref{mangano}]
and can then be numerically or analytically integrated. 

Nowadays, 
also next-to-leading order (NLO) calculations are
feasible in a direct manner, as witnessed by the accelerated
production rate of new NLO computations [\ref{lp97}].
This achievement is the result of much theoretical progress in the last few
years.
This progress regards both efficient techniques for the evaluation of one-loop
matrix elements [\ref{BDKrev}] and completely general algorithms 
[\ref{GG}--\ref{CSdipole}] to handle and cancel infrared singularities when
combining tree-level and one-loop contributions in the evaluation of physical
quantities.

In contrast,
the calculation of jet observables at next-to-next-to-leading order (NNLO),
although feasible in principle, remains a challenge in practice. Indeed,
no computations have appeared that involve more than two kinematic variables
[\ref{vanneer}]. The general obstacles that have to be overcome to perform NNLO
calculations are the evaluation of two-loop amplitudes and the cancellation
of infrared singularities between real and virtual contributions. 
There is
only one QCD amplitude that is known at two-loop order, namely the
electromagnetic form factor of the quark [\ref{gonsalves}, \ref{matsuura}].
The extension to higher loops of cutting techniques [\ref{BDKrev}] has recently
led to the computation of the two-loop four-gluon amplitude in the simplified
case of $N=4$ supersymmetric QCD [\ref{bern}]. Other approaches for
efficiently calculating higher-loop amplitudes taken by various authors include
string theory [\ref{magnea}], first quantized [\ref{fqa}] and recursive 
[\ref{reca}] approaches. Regarding the cancellation of infrared singularities,
much work still has to be done [\ref{aude}]
to extend the NLO algorithms to higher orders.

In the case of calculations at NLO, an important ingredient of the theoretical
progress mentioned above has been the complete understanding of the 
factorization properties of tree-level [\ref{book}] and one-loop
[\ref{KST}, \ref{GG}, \ref{CSdipole}] amplitudes in the
soft and collinear limits. An analogous understanding at NNLO is important as
well. In the case of tree-level amplitudes, the factorization structure
in the double-soft [\ref{dsoft}] and double-collinear [\ref{glover}] limits is 
known. The collinear limit of one-loop amplitudes [\ref{1loopcol}] is also 
known. The singular behaviour of two-loop amplitudes has not been 
investigated so far. 

In this paper we discuss the structure of 
infrared singularities in on-shell QCD amplitudes at two-loop order and,
in particular, we present a universal factorization formula that gives
the coefficients of the $\ep$-poles of the dimensionally regularized
divergences. Our results have been obtained by improving and extending the
coherent-state approach of Ref.~[\ref{CC}] to higher infrared accuracy.
More details on the derivation of the results presented here and 
on the singular behaviour of {\em all-loop} amplitudes will be given 
elsewhere.

The outline of the paper is as follows. In Sect.~\ref{notat} we introduce our
notation and, in particular, we point out the subtleties related to the
regularization scheme of the infrared divergences  
in the general context of dimensional regularization. Section \ref{sing1loop}
is devoted to a review of the known results on the singular behaviour 
of one-loop amplitudes.
Our general factorization formula for two-loop amplitudes is presented
and discussed in Sect.~\ref{sing2loop} and applications to some particular
cases are considered in Sects.~\ref{2partonsec} and \ref{3partonsec}. At
present, we are unable to explicitly compute the coefficient of the single 
poles in general terms, but we can do that in the particular case of amplitudes
with two quarks, as shown in Sect.~\ref{qqbarsec}. 
Our results are summarized in Sect.~\ref{sum}.

\section{Notation}
\label{notat}

\subsection{Dimensional regularization}
\label{dimreg}

In the evaluation of loop amplitudes one encounters ultraviolet and infrared
singularities that have to be properly regularized. The most efficient
method to simultaneously
regularize both singularities in gauge theories is to use dimensional
regularization [\ref{tHV}--\ref{cdrir}].

The key ingredient of dimensional regularization is the 
analytic continuation of loop momenta to $d=4-2\ep$ space-time dimensions.
Having done this, one is left with some
freedom regarding the dimensionality of the momenta of the external
particles as well as the number of polarizations of both external and
internal particles. This leads to different regularization schemes (RS)
within the dimensional-regularization prescription.

The RS known as conventional dimensional regularization ($\CDR$) is, in a sense,
the most natural scheme because
no distinction is made between particles in the loops and external particles.
All particle momenta are $d$-dimensional and one considers 
$d-2$ helicity states for gluons and 2
helicity states for massless quarks.
Nonetheless, other RS~\footnote{The RS that are mostly used in one-loop 
computations are the 't~Hooft and Veltman scheme [\ref{tHV}], the 
dimensional-reduction scheme
[\ref{Sie79}] and the four-dimensional helicity scheme [\ref{4dhs}].}
are known to be useful for field theoretical reasons
(e.g. to explicitly preserve 
supersymmetric Ward identities [\ref{Sie79}]) and/or because of
practical simplifications in actual calculations [\ref{BDKrev}].

This brief overview of dimensional regularization is important
to discuss the singular behaviour of loop amplitudes, because only the leading
singularities are RS-independent. In general, the coefficients of all the other
singular terms as well as the finite parts do depend on the RS. The RS
dependence regards both ultraviolet and infrared singularities, but it
affects them in different ways.

In principle the regularized ultraviolet divergences have to be removed
from off-shell Green functions via renormalization of wave functions and
coupling constant. This leads to the introduction of the running coupling
$\as(\mu^2)$, whose definition in terms of the bare coupling
depends on the regularization (and renormalization) scheme. The ensuing
scheme dependence in renormalized on-shell amplitudes is nonetheless trivial.
It simply amounts to (and can be controlled by) an overall perturbative shift 
in $\as$ (cf. Sect.~\ref{ren}). 
This perturbative shift has no effect on tree-amplitudes
and is universal in loop amplitudes, that is, independent of the
number and type of particles.

Infrared singularities behave differently with respect to 
dimensional-regularization prescriptions.
Although one can keep track of their RS dependence in
general terms [\ref{uni}], to relate loop amplitudes in different RS
one has to introduce proper transition factors that depend on the number 
and type of external partons. In the context of one-loop amplitudes,
this feature was pointed out in Ref.~[\ref{KST2to2}], where the (one-loop)
transition factors between the main schemes in current usage were 
explicitly computed. Beyond one-loop order, care has to be taken in the
operational definition of some RS. We do not explicitly consider this problem, 
but in Sects.~\ref{sing2loop}--\ref{3partonsec} 
we discuss the RS dependence of two-loop amplitudes
within the general class of schemes specified at the end of Sect.~\ref{ren}.

Note also that, in order to fulfil unitarity in
calculations of physical quantities, the RS dependence of loop amplitudes
has to be consistently matched to that of tree amplitudes. This issue
is discussed on quite a general basis in Ref.~[\ref{uni}] and is no longer
considered in this paper.

\subsection{Renormalized amplitudes}
\label{ren}

We consider amplitudes $\cm_m$ that involve $m$ external QCD partons (gluons
and massless quarks) with momenta $p_1, \dots, p_m$ and an arbitrary number
and type of particles with no colour (photons, leptons, vector bosons, ...).
Note that, by definition, we always consider incoming and outgoing parton 
momenta in the physical region, i.e. any $p_i$ is massless and with 
positive-definite energy (in particular, $p_i\cdot p_j > 0$).
The amplitudes are denoted by $\cm_m(p_1, \dots, p_m)$ (or, shortly,
$\cm_m(\{p\})$) and the dependence on the momenta and quantum numbers of 
non-QCD particles is always understood. Besides that, the unrenormalized
amplitude $\cm_m$ depends on powers of $\as^u \;\mu_0^{2\ep}$, where
$\as^u$ is the bare coupling constant and $\mu_0$ is the
dimensional-regularization scale.

We find it convenient to use renormalized amplitudes. These are
obtained from the unrenormalized ones by just expressing the bare coupling 
$\as^u$ in terms of the running coupling $\as(\mu^2)$ evaluated at the 
arbitrary renormalization scale $\mu^2$. In order to avoid 
renormalization-scheme ambiguities, we always consider the running coupling
$\as(\mu^2)$ as that defined in the standard ${\overline {\rm MS}}$ scheme. 
Thus, we use the following expression
\beq
\label{mscoup}
\as^u \;\mu_0^{2\ep} \;S_{\ep} = \as(\mu^2) \;\mu^{2\ep} 
\left[ 1 -  \as(\mu^2) \;\frac{\beta_0}{\ep} + \as^2(\mu^2)
\left( \frac{\beta_0^2}{\ep^2} - \frac{\beta_1}{2 \ep} \right) 
+ {\cal O}(\as^3(\mu^2)) \right] \;\;,
\eeq 
where $\beta_0, \beta_1$ are the first two coefficients of the QCD beta
function
\beq
\label{betas}
\beta_0 = \frac{11 C_A - 4 T_R N_f}{12\pi} \;\;, \;\; \;\;\;\;
\beta_1 = \frac{17 C_A^2 - 10 C_A T_R N_f - 6 C_F T_R N_f}{24 \pi^2} \;\;,
\eeq
and $S_{\ep}$ is the typical phase-space volume factor in $d=4-2\ep$
dimensions ($\gamma_E= - \psi(1) = 0.5772\dots$ is the Euler number)
\beq
\label{spher}
S_{\ep} = \exp \left[ \,\ep \,(\ln 4\pi + \psi(1)) \right] \;\;.
\eeq

Note that, as recalled in Sect.~\ref{dimreg}, unrenormalized amplitudes
contain some RS dependence of ultraviolet origin. Using the  
${\overline {\rm MS}}$ renormalization scheme, we have eliminated
this harmless dependence from renormalized amplitudes. In practical terms,
our renormalization procedure works as follows. If the bare amplitude is
evaluated in $\CDR$, one simply replaces $\as^u$ according to 
Eq.~(\ref{mscoup}). If a different RS is used, the minimally-subtracted
substitution in Eq.~(\ref{mscoup}) still applies in terms of an
RS-dependent coupling $\as^{\RS}(\mu^2)$. To obtain the ${\overline {\rm MS}}$ 
amplitudes, one then has to perform the additional substitution:  
\beq
\label{uvsub}
\as^{\RS} = \as \left[ 1 + d_1 \,\as + d_2 \,\as^2 + {\cal O}(\as^3) \right] 
\;\;,
\eeq
where the coefficients $d_1, d_2$ depend on the RS and are finite for 
$\ep \to 0$.

The ${\overline {\rm MS}}$-renormalized amplitude has the following 
perturbative expansion:
\beeq
\label{loopex}
\cm_m(\as(\mu^2), \mu^2; \{p\}) = \left( \frac{\as(\mu^2)}{2\pi} \right)^q
&\!&\!\!\! \!\!\! \!\!\! \!\!\! \left[ \; \cm_m^{(0)}(\mu^2; \{p\}) 
+ \frac{\as(\mu^2)}{2\pi} \;\cm_m^{(1)}(\mu^2; \{p\}) \right. \nonumber \\
&+& \!\!
\left. \left( \frac{\as(\mu^2)}{2\pi} \right)^2 \;\cm_m^{(2)}(\mu^2; \{p\})
+ {\cal O}(\as^3(\mu^2)) \right] \;\;,
\eeeq
where the overall power $q$ is half-integer ($q=0,1/2,1,3/2,\dots$), in general.
This equation fixes the normalization of the tree-level $(\cm_m^{(0)})$,
one-loop $(\cm_m^{(1)})$ and two-loop $(\cm_m^{(2)})$ coefficient
amplitudes that we use in the rest of the paper\footnote{Precisely
speaking, $\cm_m^{(0)}$ is not necessarily a tree amplitude, but rather the
lowest-order amplitude for that given process; $\cm_m^{(1)}, \cm_m^{(2)}$
are the corresponding one-loop and two-loop corrections. For instance, in the
case of $gg \to \gamma \gamma$, $\cm^{(0)}$ involves a quark loop.}. 
These coefficient
amplitudes are infrared-singular when $\ep \to 0$ and behave as
\beq
\label{singbeh}
\cm_m^{(n)} \sim \left( \frac{1}{\ep} \right)^{2n} + \dots \;\;,
\eeq
where the dots stand for $\ep$-poles of lower order. The coefficient of
the poles are still RS-dependent. The results presented in the following
sections are valid in $\CDR$ and in any other RS that is consistently defined 
as ultraviolet regulator. By that, we mean in any other scheme whose 
minimally-subtracted coupling 
is related to the ${\overline {\rm MS}}$ coupling via Eq.~(\ref{uvsub}) in 
a universal way (i.e. with coefficients $d_1, d_2$ that do not depend on the
computed amplitude).

\subsection{Colour space}
\label{col}

The colour structure of QCD amplitudes can be handled in two different
and equivalent ways. One method consists in projecting out the amplitude
onto a particular set of non-orthogonal colour vectors, thus extracting 
colour subamplitudes [\ref{mangano}]. The other method works directly in colour
space. We prefer to use the colour-space formalism and, in particular,  
the same notation as in Ref.~[\ref{CSdipole}].

The colour indices of the $m$ partons in the amplitude $\cm_m$ are generically
denoted by $c_1,\dots,c_m$: $c_i=\{a \}= 1,\dots,N_c^2-1$ for gluons and
$c_i=\{ \alpha \}=1,\dots,N_c$ for quarks and antiquarks.
We formally introduce an orthogonal basis of unit
vectors $\{ |c_1,\dots,c_m \ra \}$ in the $m$-parton colour space,
in such a way that the colour amplitude can be written
as follows:
\beq
\label{medef}
\cm_m^{c_1,\dots,c_m}(p_1,\dots,p_m) \equiv
\la c_1,\dots,c_n \,
| \, \cm_m(p_1,\ldots,p_m)\ra \;\;.
\eeq
Thus
$|\cm_m(p_1,\ldots,p_m)\ra$ is an abstract vector in colour space and,
in particular, the square amplitude summed over colours is:
\beq
|\cm_m(\{p\})|^2 = 
\la \cm_m(\{p\}) \, | \, \cm_m(\{p\}) \ra \;\;.
\eeq

Colour interactions at the QCD vertices are represented by associating
a colour charge ${\bf T}_i$ with the emission of a gluon from each parton $i$.
The colour charge ${\bf T}_i= \{T_i^a \} $ is a vector with respect to the 
colour indices $a$ of the emitted gluon and an 
$SU(N_c)$ matrix with respect to the colour indices of the parton $i$.
More precisely, its action onto the colour space is defined by
\beq
\la c_1,\dots,c_i,\dots,c_m \,| \, T_i^a \, | \,b_1,\dots,b_i,\dots,b_m \ra 
= \delta_{c_1b_1} \dots T_{c_ib_i}^a \dots \delta_{c_mb_m} \;,
\eeq
where $T_{c b}^a \equiv i f_{cab}$ (colour-charge matrix
in the adjoint representation) if the emitting particle $i$ 
is a gluon and $T_{\alpha \beta}^a \equiv t^a_{\alpha \beta}$
(colour-charge matrix in the fundamental representation) 
if the emitting particle $i$ is a final-state quark (in the case of a
final-state antiquark $T_{\alpha \beta}^a \equiv {\bar t}^a_{\alpha \beta}
= - t^a_{\beta \alpha }$). Note that the colour-charge operator of an
initial-state parton is defined by crossing symmetry, that is by
$({\bf T}_i)^a_{\alpha \beta} = {\bar t}^a_{\alpha \beta}
= - t^a_{\beta \alpha }$ if $i$ is an initial-state quark and
$({\bf T}_i)^a_{\alpha \beta} = t^a_{\beta \alpha }$ if $i$ is an initial-state
antiquark.

In this notation, each vector $|\cm_m(p_1, ..., p_m) \ra$ is a colour singlet, 
so colour conservation is simply
\beq
\label{colcon}
\sum_{i=1}^m {\bom T}_i \;|\cm_m \ra = 0 \;.
\eeq

The colour-charge algebra for the product 
$({\bf T}_i)^a ({\bf T}_j)^a \equiv {\bf T}_i \cdot {\bf T}_j$ is:
\beq
{\bom T}_i \cdot {\bom T}_j ={\bom T}_j \cdot {\bom T}_i \;\;\;\;{\rm if}
\;\;i \neq j; \;\;\;\;\;\;{\bom T}_i^2= C_i \;,
\eeq
where $C_i$ is the Casimir operator, and we have
$C_i=C_A=N_c$ if $i$ is a gluon and $C_i=C_F= T_R(N_c^2-1)/N_c=
(N_c^2-1)/2N_c$ if $i$ is a quark or an antiquark (we are using the customary
normalization $T_R=1/2$).
 
\section{Singular behaviour at one-loop order}
\label{sing1loop}

In this section we recall the known results 
[\ref{KST}, \ref{GG}, \ref{CSdipole}]
on the infrared-singular behaviour of QCD amplitudes at one-loop order. The
one-loop coefficient subamplitude $\cm_m^{(1)}(\mu^2; \{p\})$ has double
and single poles in $1/\ep$. The coefficients of these poles are universal and
can be given by the following formula
\beq
\label{ff1loop}
| \cm_m^{(1)}(\mu^2;\{p\}) \ra_{\RS} = {\bom I}^{(1)}(\ep,\mu^2;\{p\}) 
\; | \cm_m^{(0)}(\mu^2;\{p\}) \ra_{\RS}
+ | \cm_m^{(1)\, {\rm fin}}(\mu^2;\{p\}) \ra_{\RS} \;\;.
\eeq

The contribution $\cm_m^{(1)\, {\rm fin}}$ on the right-hand side 
is finite for $\ep \to 0$ and, hence, in Eq.~(\ref{ff1loop})
all one-loop singularities are factorized in colour
space with respect to the tree-level amplitude $\cm_m^{(0)}$. The singular
dependence is embodied in the factor ${\bom I}^{(1)}$ that acts as a
colour-charge operator onto the colour vector $| \cm_m^{(0)} \ra$. Its
explicit expression in terms of the colour charges of the $m$ partons is 
\beq
\label{iee}
{\bom I}^{(1)}(\ep,\mu^2;\{p\}) =  \frac{1}{2}
\frac{e^{-\ep \psi(1)}}{\Gamma(1-\ep)} \sum_i \;\frac{1}{{\bom T}_{i}^2} \;{\cal
V}_i^{{\rm sing}}(\ep)
\; \sum_{j \neq i} {\bom T}_i \cdot {\bom T}_j
\; \left( \frac{\mu^2 \phase}{2 p_i\cdot p_j} \right)^{\ep} \;\;,
\eeq
where $\phase$ is the unitarity phase ($\lambda_{ij}=+1$ if $i$ and $j$
are both incoming or outgoing partons and $\lambda_{ij}=0$ otherwise) 
and the singular (for
$\ep \to 0$) function ${\cal V}_i^{{\rm sing}}(\ep)$ depends only on the 
parton flavour and is given by
\beq
\label{calvexp}
{\cal V}_i^{{\rm sing}}(\ep) = {\bom T}_{i}^2 \frac{1}{\ep^2} 
+ \gamma_i \;\frac{1}{\ep} \;\;.
\eeq
The flavour coefficients ${\bom T}_{i}^2$ and $\gamma_i$ are
\beeq
{\bom T}_q^2 = {\bom T}_{\bar q}^2 = C_F  \;\;, 
\;\;\;\;\;\;&&{\bom T}_g^2 = C_A \;\;, \nonumber \\
\gamma_q = \gamma_{\bar q} = \frac{3}{2} \,C_F \;\;, 
\;\;\;\;\;\;&&\gamma_g = \frac{11}{6} \, C_A -  \frac{2}{3} \,T_R N_f \;\;.
\eeeq

Note that in Eq.~(\ref{ff1loop}) the double poles $1/\ep^2$ are 
factorized completely and not only in colour space. The simplest way to see
that is to expand Eq.~(\ref{iee}) in powers of $\ep$ and then use the colour
conservation relation (\ref{colcon}), 
i.e. $\sum_{j \neq i} {\bom T}_j = - {\bom T}_i$. One obtains the result
\beq
\label{ieeexp}
{\bom I}^{(1)}(\ep,\mu^2;\{p\}) =  \frac{1}{2}
\sum_i \frac{1}{\ep^2} \; \sum_{j \neq i} {\bom T}_i \cdot {\bom T}_j
+ {\cal O}(1/\ep) = - \frac{1}{2\ep^2} \sum_i {\bom T}_i^2
+ {\cal O}(1/\ep) \;\;,
\eeq
that explicitly shows the absence of colour correlations at ${\cal O}(1/\ep^2)$.
Nonetheless, single poles $1/\ep$ are still colour-correlated.

In the factorization formula
(\ref{ff1loop}) we have introduced the subscripts $\RS$
to explicitly recall the RS dependence of the various quantities.
Note that the insertion operator ${\bom I}^{(1)}$ is RS-independent.
The RS dependence of the one-loop amplitude $\cm_m^{(1)}$ is completely taken
into account by that of the tree amplitude $\cm_m^{(0)}$ and by the
finite remainder $\cm_m^{(1)\;{\rm fin}}$. In particular, the interference
between RS-dependent terms\footnote{Note
that the $\ep$-dependence of any amplitude $\cm_m$ is always understood.}
of ${\cal O}(\ep)$ in $\cm_m^{(0)}$ and double poles
$1/\ep^2$ in ${\bom I}^{(1)}$ produces, in general, 
an RS dependence of $\cm_m^{(1)}$ that begins at ${\cal O}(1/\ep)$. 

The RS-dependent terms of $\cm_m^{(1)\;{\rm fin}}$ are not 
colour-correlated. Since they are completely factorized, we can avoid the
colour-space notation and write
\beq
\label{fin1loop}
\cm_{m, \, \RS}^{(1)\, {\rm fin}}(\mu^2;\{p\}) = \frac{1}{2}
\left( \sum_i {\tilde \gamma}_i^{\RS} \right)
\cm_{m, \, \RS}^{(0)}(\mu^2;\{p\}) + F_m^{(1)}(\mu^2;\{p\}) + 
{\cal O}(\ep)\;\;,
\eeq
where the finite  coefficients ${\tilde \gamma}_i^{\RS}$ 
depend only on the flavour of the external partons in  $\cm_m$.
The transition coefficients ${\tilde \gamma}_i^{\RS}$ that relate
the RS mostly used in one-loop computations were first calculated in
Ref.~[\ref{KST2to2}]. Using the normalization in Eq.~(\ref{fin1loop})
and assuming $\CDR$ as reference scheme 
(i.e. setting ${\tilde \gamma}_i^{\cdr}=0$, by definition), the explicit 
expressions of ${\tilde \gamma}_i^{\RS}$ can be found in 
Ref.~[\ref{uni}].

The residual RS dependence on the right-hand side of Eq.~(\ref{fin1loop})
affects only the terms of ${\cal O}(\ep)$.
In particular, the $\ep$-independent function $F_m^{(1)}(\mu^2;\{p\})$
is RS-independent [\ref{KST2to2}, \ref{uni}].

\section{Factorization formula at two-loop order}
\label{sing2loop}

The two-loop coefficient subamplitude $\cm_m^{(2)}$ has $1/\ep^4, 1/\ep^3,
1/\ep^2$ and $1/\ep$ poles. Because of the increased degree of singularities,
it is not a priori guaranteed that all of them can be controlled by a {\em
universal} factorization formula as in the one-loop case. The  
main result
presented in this paper is that such a factorization formula does exist and can
be written in the following form
\beeq
\label{ff2loop'}
| \cm_m^{(2)}(\mu^2; \{p\}) \ra_{\RS} &=& {\bom I}^{(1)}(\ep, \mu^2; \{p\}) 
\; | \cm_m^{(1)}(\mu^2; \{p\}) \ra_{\RS} \nonumber \\ \label{ff2loop}
&+& 
{\bom I}^{(2)}_{\RS}(\ep, \mu^2; \{p\}) \; | \cm_m^{(0)}(\mu^2; \{p\}) \ra_{\RS}
+ \; | \cm_m^{(2) {\rm fin}}(\mu^2; \{p\}) \ra_{\RS} \;\;.
\eeeq
The factorization structure of Eq.~(\ref{ff2loop}) is not trivial and
is somehow analogous to that of the 
collinear limit of one-loop amplitudes considered in Ref.~[\ref{1loopcol}].

The main features of Eq.~(\ref{ff2loop}) are the following:
\begin{itemize}
\item
In the first term on the right-hand side of Eq.~(\ref{ff2loop}), the one-loop
insertion operator ${\bom I}^{(1)}$ of Eq.~(\ref{iee}) acts onto the one-loop
matrix element. The corresponding contribution thus contains poles of the
type $1/\ep^n$ with $n=1,\dots,4$, coming from the single and double poles in
${\bom I}^{(1)}$ and $\cm_m^{(1)}$. 

\item
The second term on the right-hand side of Eq.~(\ref{ff2loop}) contains a new
colour-charge operator
${\bom I}^{(2)}$ that acts onto the tree-level subamplitude. 
The two-loop insertion operator ${\bom I}^{(2)}$ is given as follows
\beeq
{\bom I}^{(2)}_{\RS}(\ep,\mu^2;\{p\}) 
&=& - \frac{1}{2} {\bom I}^{(1)}(\ep,\mu^2;\{p\})
\left( {\bom I}^{(1)}(\ep,\mu^2;\{p\}) + 4 \pi \beta_0 \frac{1}{\ep} \right)
\nonumber \\
&+& \frac{e^{+\ep \psi(1)} \Gamma(1-2\ep)}{\Gamma(1-\ep)} 
\left(2 \pi \beta_0 \frac{1}{\ep} + K \right) {\bom I}^{(1)}(2\ep,\mu^2;\{p\})
\label{iop2loop}
\\
&+& {\bom H}^{(2)}_{\RS}(\ep,\mu^2;\{p\}) \nonumber \;\;,
\eeeq
where the coefficient $K$ is:
\beq
\label{kcoef}
K = \left( \frac{67}{18} - \frac{\pi^2}{6} \right) C_A - \frac{10}{9} T_R N_f
\;\;.
\eeq
The first and second line\footnote{Note that on the second line the argument
of the operator ${\bom I}^{(1)}$ is $2 \ep$ rather than $\ep$ as on the first
line.} on the right-hand side of Eq.~(\ref{iop2loop}) lead to $\ep$ poles
that are at most of 4th- and 3rd-order, respectively. Moreover, these two
lines control all the poles up to $1/\ep^2$. In fact, the remaining term
${\bom H}^{(2)}_{\RS}$ contains only {\em single} poles: 
\beq
\label{h21ep}
{\bom H}^{(2)}_{\RS}(\ep,\mu^2;\{p\}) = {\cal O}(1/\ep) \;\;.
\eeq

\item
The last term, $\cm_m^{(2) {\rm fin}}$, on the right-hand side of 
Eq.~(\ref{ff2loop}) is a non-singular remainder in the limit $\ep \to 0$.
\end{itemize}

Note that 
the two-loop operator ${\bom I}^{(2)}_{\RS}$ in Eq.~(\ref{iop2loop}) depends
on the RS only through the single-pole contributions in ${\bom H}^{(2)}_{\RS}$.
At present, we cannot give an explicit expression for ${\bom H}^{(2)}_{\RS}$
that is valid for any QCD amplitude. The particular case of a $q{\bar q}$ pair
is considered in Sect.~\ref{qqbarsec}. 

Using the factorization formula (\ref{ff2loop}) and Eq.~(\ref{iop2loop}), 
all the coefficients of the poles $1/\ep^4, 1/\ep^3, 1/\ep^2$ can be explicitly
evaluated in terms of the one-loop operator ${\bom I}^{(1)}$, the first
coefficient $\beta_0$ of the beta function and the constant $K$ in
Eq.~(\ref{kcoef}). 

In particular, the coefficient $K$ allows one to
control the double poles $1/\ep^2$ in a universal way, that is, independently
of the given QCD amplitude. The same coefficient typically appears in the
resummation program of higher-order logarithmic corrections of the Sudakov type
[\ref{Kfactor}]. This {\em universality}
[\ref{CMW}] follows from the fact 
that $K$ measures the renormalization (in the ${\overline {\rm MS}}$ 
renormalization scheme) of the intensity of the lowest-order soft-gluon 
emission. The origin of $K$ is thus essentially ultraviolet, which explains
its independence of the RS of infrared singularities.  

The non-singular remainder $\cm_m^{(2) {\rm fin}}$ is anologous to 
$\cm_m^{(1) {\rm fin}}$ in Eq.~(\ref{ff1loop}). However, 
its RS-dependent part is in general not factorized in colour space and
cannot be expressed in factorized form as in
Eq.~(\ref{fin1loop}).

\section{Two-loop amplitudes with two partons}
\label{2partonsec}

\subsection{Colour structure}

Amplitudes with only two QCD partons
have a trivial colour structure to any loop order. The colour-space 
formulae of Sects.~\ref{sing1loop} and \ref{sing2loop}
thus become `true' factorization formulae because all
the insertion operators are proportional to the identity matrix in colour space.
  
In the case of the one-loop insertion operator ${\bom I}^{(1)}$, one can use
colour conservation as in Eq.~(\ref{colcon}) and explicitly perform the colour
algebra, that is 
${\bom T}_1 \cdot {\bom T}_2 = - {\bom T}_1^2 = - {\bom T}_2^2$. Then one
obtains
\beq
\label{iop2}
{\bom I}^{(1)}_{ij}(\ep,\mu^2;p_1,p_2) = - {\cal V}_i^{{\rm sing}}(\ep)
\; \frac{e^{-\ep \psi(1)}}{\Gamma(1-\ep)}
\; \left( \frac{\mu^2 \phasot}{2 p_1\cdot p_2} \right)^{\ep} \;\;,
\eeq
where the subscript $ij$ 
denotes the flavour of the
two partons: $ij=gg$ or $ij=q{\bar q}, qq, {\bar q}{\bar q}$ for the various
crossed channels with two fermions. Inserting Eq.~(\ref{iop2}) into
Eq.~(\ref{iop2loop}), one can thus rewrite Eq.~(\ref{ff2loop}) in an explicitly
factorized form.

\subsection{Amplitudes with a $q {\bar q}$ pair:
complete structure of singular terms in CDR}
\label{qqbarsec}

The only QCD amplitude that has been computed so far at two-loop order is
the electromagnetic form factor of the quark [\ref{gonsalves}, \ref{matsuura}].
The result of Ref.~[\ref{matsuura}] and Eq.~(\ref{iop2}) can be used to check 
our Eqs.~(\ref{ff2loop}) and (\ref{iop2loop}) in this particular case. 
As a by-product of
this check, we can moreover derive the explicit expression of the
${\cal O}(1/\ep)$ operator ${\bom H}_{q{\bar q}}^{(2)}$ for the
QCD amplitudes with a $q{\bar q}$ pair or, in general, for those related to
them by crossing symmetry.   
We obtain
\beeq
\label{H2qqbar}
{\bom H}^{(2)}_{q{\bar q}, \,\cdr}(\ep, \mu^2; p_1,p_2) &=& \frac{1}{4\ep}
\; \frac{e^{-\ep \psi(1)}}{\Gamma(1-\ep)} 
\;\left( \frac{\mu^2 \phasot}{2 p_1\cdot p_2} \right)^{2\ep} \\
&\cdot&
\left[ \frac{1}{4} \;\gamma_{(1)} + 3 C_F K + 5 \zeta_2 \pi \beta_0 
C_F -\frac{56}{9} \pi \beta_0 C_F - \left(\frac{16}{9} - 7 \zeta_3 \right)
C_F C_A \right] \;, \nonumber
\eeeq
where
\beq
\gamma_{(1)} = (-3 +24 \zeta_2 - 48 \zeta_3) C_F^2 
+ \left(-\frac{17}{3} - \frac{88}{3} \zeta_2 + 24 \zeta_3 \right) C_F C_A
+ \left(\frac{4}{3} + \frac{32}{3} \zeta_2 \right)C_F T_R N_f \;.
\eeq
Note that the $C_F^2$-part of ${\bom H}^{(2)}_{q{\bar q}}$ ie entirely
included in the coefficient $\gamma_{(1)}$. This is exactly the same
coefficient as controls the virtual contribution (i.e. the term proportional
to $\delta(1-z)$) to the NLO Altarelli--Parisi splitting function in the
flavour non-singlet sector [\ref{CFP}].

We recall that the operator in Eq.~(\ref{H2qqbar}) is not only related
to the electromagnetic form factor of the quark. The same operator applies
to the two-loop factorization formula of 
any QCD amplitude with two quarks such as, for instance, the amplitude 
$q {\bar q} \to \gamma \gamma$, which is relevant for the hadroproduction of  
diphotons. Note also that expression (\ref{H2qqbar}) is valid only in CDR
because it is obtained by a calculation [\ref{matsuura}] that uses
this RS.

\section{Two-loop amplitudes with three partons}
\label{3partonsec}

\subsection{Colour structure}

In the case of QCD amplitudes with three partons, the one-loop insertion
operator ${\bom I}^{(1)}$ is again factorizable in colour space. Using colour
conservation (i.e. Eq.~(\ref{colcon})), we have 
$2 {\bom T}_1 \cdot {\bom T}_2 = {\bom T}_3^2- {\bom T}_1^2 - {\bom T}_2^2$
and likewise for the other permutations. Then, in the case of amplitudes
with three gluons, we obtain:
\beq
\label{i13g}
{\bom I}^{(1)}_{ggg}(\ep,\mu^2;p_1,p_2,p_3) \!= - \frac{1}{2} \,
{\cal V}_g^{{\rm sing}}(\ep)
\; \frac{e^{-\ep \psi(1)}}{\Gamma(1-\ep)}
\!\left[ \left( \frac{\mu^2 \phasot}{2 p_1\cdot p_2} \right)^{\ep} 
+ \left( \frac{\mu^2 \phastt}{2 p_2\cdot p_3} \right)^{\ep} 
+ \left( \frac{\mu^2 \phasto}{2 p_3\cdot p_1} \right)^{\ep} \right] ,
\eeq
and, for amplitudes with two fermions of momenta $p_1,p_2$ and a gluon of
momentum $p_3$, we get:
\beeq
\label{i1qqg}
\!\! \!\! \!\! {\bom I}^{(1)}_{q{\bar q}g}(\ep,\mu^2;p_1,p_2,p_3) 
&=& - \frac{1}{2} \,
\frac{e^{-\ep \psi(1)}}{\Gamma(1-\ep)}
\left\{ {\cal V}_q^{{\rm sing}}(\ep) \, \frac{1}{C_F} \,(2 C_F - C_A)
\left( \frac{\mu^2 \phasot}{2 p_1\cdot p_2} \right)^{\ep} \right.  \\
&+& \left. \frac{1}{2} \left( {\cal V}_g^{{\rm sing}}(\ep)
+ \frac{C_A}{C_F} \, {\cal V}_q^{{\rm sing}}(\ep) \right)
\left[ \left( \frac{\mu^2 \phastt}{2 p_2\cdot p_3} \right)^{\ep} 
+ \left( \frac{\mu^2 \phasto}{2 p_3\cdot p_1} \right)^{\ep} \right] \right\}
\;. \nonumber
\eeeq

The fact that the insertion operator ${\bom I}^{(1)}$ is proportional to the
identity matrix in colour space
 does not imply, however,
that formula (\ref{ff2loop}) is exactly factorizable 
in the three-parton case. In fact, colour correlations can still be present
in ${\bom I}^{(2)}$ through
the insertion operator ${\bom H}^{(2)}$. When acting onto the tree-level
matrix element $\cm_m^{(0)}$,
the operator ${\bom H}^{(2)}$ can lead to colour transitions between its
possible colour states. 

In this respect, $q{\bar q}g$ and $ggg$ amplitudes
behave in a different way. In the $q{\bar q}g$ case, there is only one possible
colour state, namely $\cm_{q{\bar q}g}^{\alpha {\bar \alpha} a} 
\propto t_{\alpha {\bar \alpha}}^{a},$
and thus ${\bom H}^{(2)}$ exactly factorizes. In the $ggg$ case, the gluons
can be either in the symmetric $(\cm_{ggg}^{abc} \propto f_{abc})$
or in the antisymmetric $(\cm_{ggg}^{abc} \propto d_{abc})$
colour configuration, and the colour structure of ${\bom H}^{(2)}$ has to
be properly taken into account.

\subsection{$q {\bar q}g$ amplitudes and \ee $\to 3$ jets at NNLO}
Perturbative QCD predictions at NNLO for three-jet observables in \ee\
annihilation are strongly demanded by the high experimental accuracy of the 
LEP and SLC data [\ref{lp97}]. 
These predictions require the calculation of the corresponding scattering
amplitude at two-loop order. Since the knowledge of its singularity structure 
can help to set up and to check the calculation,  
we apply in this section the factorization formula (\ref{ff2loop}) to this
particular case. 

We are interested in the process $e^+ e^- \to q(p_1) + {\bar q}(p_2) + g(p_3)$,
where $Q = p_1 + p_2 + p_3$ denotes the total four-momentum and 
$y_{ij} = 2p_i \cdot p_j/Q^2$ are the relevant dimensionless invariants.
To compute the corresponding cross section at NNLO, one has to evaluate
the interference $| \cm_{q{\bar q}g} |^2_{({\rm 2-loop})}$
between the tree-level and two-loop matrix elements:
\beq
| \cm_{q{\bar q}g} |^2_{({\rm 2-loop})} = 
\la \cm_{q{\bar q}g}^{(0)} \, | \, 
\cm_{q{\bar q}g}^{(2)} \ra \;
+ {\rm complex \; conjugate} \;\;.
\eeq 
The analogous interference
$| \cm_{q{\bar q}g} |^2_{({\rm 1-loop})}$ at the one-loop level,
\beq
| \cm_{q{\bar q}g} |^2_{({\rm 1-loop})} = 
\la \cm_{q{\bar q}g}^{(0)} \, | \, 
\cm_{q{\bar q}g}^{(1)} \ra \;
+ {\rm complex \; conjugate} \;\;,
\eeq 
was first computed in Refs.~[\ref{ERT}, \ref{kramer}].

Using the factorization formulae (\ref{ff1loop}) and (\ref{ff2loop}), 
we can write
\beq
| \cm_{q{\bar q}g} |^2_{({\rm 2-loop})} = 2 \;
| \cm_{q{\bar q}g}^{(0)} |^2 \left[ {\rm Re} \,{\bom I}^{(2)}_{q{\bar q}g}
- \left( {\rm Im} \,{\bom I}^{(1)}_{q{\bar q}g} \right)^2 \right] 
+ | \cm_{q{\bar q}g} |^2_{({\rm 1-loop})} \;
{\rm Re}\, {\bom I}^{(1)}_{q{\bar q}g} + {\cal O}(1/\ep) \;\;.
\eeq  
Then, we can use the explicit expression for 
$| \cm_{q{\bar q}g} |^2_{({\rm 1-loop})}$ and Eqs.~(\ref{i1qqg}) and
(\ref{iop2loop}) and obtain the final result: 
\beeq
| \cm_{q{\bar q}g} |^2_{({\rm 2-loop})} \!&=\!\!&\! | 
\cm_{q{\bar q}g}^{(0)} |^2_{\RS}
\left\{ \frac{1}{4\ep^4} f_{2}^2(\ep) + \frac{1}{2\ep^3}
\left[ f_{2}(\ep) \, f_{1}(\ep) + 6 \pi \beta_0 f_{2}(\ep) - 
\pi \beta_0 f_{2}(2\ep) \right] \right. \nonumber \\
\!&+\!\!&\! \left. \frac{1}{4\ep^2} \left[ f_{1}^2(\ep) - \pi^2 f_{2}^2(\ep)
+ 8 \pi \beta_0 f_{1}(\ep) + 12 \pi^2 \beta_0^2 + 2 L_{\RS} f_{2}(\ep) 
- K f_{2}(2\ep)
 \right] \right\} \nonumber \\
\label{m2loop}
\!&-\!\!&\! \frac{1}{2\ep^2} f_{2}(\ep) \;
F(y_{12},y_{23},y_{13}) + {\cal O}(1/\ep) \;\;, 
\eeeq
where the function $F(y_{12},y_{23},y_{13})$ is given in Eq.~(2.21) of
Ref.~[\ref{ERT}] and the functions $f_{2}, f_{1}$ and the coefficient $L_{\RS}$ 
are defined as follows
\beq
f_{2}(\ep) = \frac{e^{-\ep \psi(1)}}{\Gamma(1-\ep)}
\left( \frac{\mu^2}{Q^2} \right)^{\ep}  
\left[ (2 C_F - C_A ) y_{12}^{-\ep} + C_A ( y_{13}^{-\ep} + y_{23}^{-\ep})
\right] \;,
\eeq
\beq
f_{1}(\ep) = \frac{e^{-\ep \psi(1)}}{\Gamma(1-\ep)}
\left( \frac{\mu^2}{Q^2} \right)^{\ep}  
3 C_F  \;,
\eeq
\beq
\label{lfun}
L_{\RS} =   - \frac{\pi^2}{2} (2 C_F + C_A) + 8 C_F
-2 {\tilde \gamma}_q^{\RS} - {\tilde \gamma}_g^{\RS} \,.
\eeq
The coefficients ${\tilde \gamma}_i^{\RS}$ in Eq.~(\ref{lfun}) parametrize the 
RS-dependence of the amplitude (see Eq.~(\ref{fin1loop})). The complete
evaluation of the ${\cal O}(1/\ep)$-terms on the right-hand side of 
Eq.~(\ref{m2loop}) requires the identification of the still unknown operator
${\bom H}^{(2)}_{q{\bar q}g}$. 

\section{Summary}
\label{sum}

In this paper we have presented a first discussion of the singular behaviour
of on-shell QCD amplitudes at two-loop order. The complete structure of
the infrared singularities is described by the colour-space factorization
formula given in Sect.~\ref{sing2loop}. The factorization formula is universal,
i.e.
valid for any amplitude, and the singular factors only depend on the flavour and
momentum of the coloured external legs. At present we can explicitly give
only the coefficients of the $1/\ep^4, 1/\ep^3$ and $1/\ep^2$ poles of the
dimensionally-regularized singular factors. These coefficients are nonetheless
known in a form that is manifestly independent of the RS of the infrared
singularities. The remaining single-pole contributions, namely the operator
${\bom H}^{(2)}$ in Eq.~(\ref{iop2loop}), still have to be explicitly evaluated.
Owing to their universality, they can be extracted from the calculation of few
basic two-loop amplitudes, as discussed in Sect.~\ref{qqbarsec} for the case 
of amplitudes with a $q{\bar q}$ pair.

Our factorization formula can be useful both to check explicit evaluations of
two-loop amplitudes and to organize their calculations in terms of divergent,
but analytically computable, parts and finite remainders that can be integrated
numerically. In the more general context of NNLO calculations of jet
observables, our two-loop results can be used to set up the integration of
tree-level and one-loop amplitudes in such a way as to construct
process-independent techniques for infrared cancellations.

\noindent {\bf Acknowledgements}. We would like to thank David Kosower,
Lorenzo Magnea and Willy van Neerven for discussions.

\section*{References}

\def\ac#1#2#3{Acta Phys.\ Polon.\ #1 (19#3) #2}
\def\ap#1#2#3{Ann.\ Phys.\ (NY) #1 (19#3) #2}
\def\ar#1#2#3{Annu.\ Rev.\ Nucl.\ Part.\ Sci.\ #1 (19#3) #2}
\def\cpc#1#2#3{Computer Phys.\ Comm.\ #1 (19#3) #2}
\def\ib#1#2#3{ibid.\ #1 (19#3) #2}
\def\np#1#2#3{Nucl.\ Phys.\ B#1 (19#3) #2}
\def\pl#1#2#3{Phys.\ Lett.\ #1B (19#3) #2}
\def\pr#1#2#3{Phys.\ Rev.\ D #1 (19#3) #2}
\def\prep#1#2#3{Phys.\ Rep.\ #1 (19#3) #2}
\def\prl#1#2#3{Phys.\ Rev.\ Lett.\ #1 (19#3) #2}
\def\rmp#1#2#3{Rev.\ Mod.\ Phys.\ #1 (19#3) #2}
\def\sj#1#2#3{Sov.\ J.\ Nucl.\ Phys.\ #1 (19#3) #2}
\def\zp#1#2#3{Z.\ Phys.\ C#1 (19#3) #2}

\begin{enumerate}

\item \label{book}
R.K.\ Ellis, W.J.\ Stirling and B.R.\ Webber, {\it QCD and collider 
physics} (Cambridge University Press, Cambridge, 1996) and references therein.

\item \label{mangano}
M.L.\ Mangano and S.J.\ Parke, \prep{200}{301}{91}
and references therein.

\item \label{lp97}
S.\ Catani, preprint CERN-TH/97-37 (hep-ph/9712442), to appear in 
Proceedings of the {\it XVIII International Symposium on Lepton-Photon
Interactions}, LP97, Hamburg (1997) and references therein.

\item \label{BDKrev}
Z.\ Bern, L.\ Dixon and D.A.\ Kosower, \ar{46}{109}{96} and references therein.

\item \label{GG}
W.T. Giele and E.W.N. Glover, \pr{46}{1980}{92}.

\item \label{GGK}
W.T. Giele, E.W.N. Glover and D.A. Kosower, \np{403}{633}{93}.

\item \label{submeth}
D.E.\ Soper and Z.\ Kunszt, \pr{46}{192}{92};
S.\ Frixione, Z.\ Kunszt and A.\ Signer, \np{467}{399}{96};
Z. Nagy and Z. Tr\'ocs\'anyi, \np{486}{189}{97};
S.\ Frixione, \np{507}{295}{97}.

\item \label{CSdipole}
S.\ Catani and M.H.\ Seymour, \pl{378}{287}{96}, \np{485}{291}{97}
(E ibid. B510 (1998) 503).

\item \label{vanneer}
R.\ Hamberg, W.L.\ van Neerven and T.\ Matsuura, \np{359}{343}{91};
E.B.~Zijlstra and W.L. van Neerven, \np{383}{525}{92};
P.J.\ Rijken and W.L.\ van Neerven, \np{487}{233}{97}.

\item \label{gonsalves}
R.J.\ Gonsalves, \pr{28}{1542}{83}.
 
\item \label{matsuura}
G.\ Kramer and B.\ Lampe, \zp{34}{497}{87} (E ibid. C42 (1989) 504);
T.~Matsuura and W.L.\ van Neerven, \zp{38}{623}{88};
T.\ Matsuura, S.C.\ van der Marck and W.L.\ van Neerven, \np{319}{570}{89}.

\item \label{bern}
Z.\ Bern, J.S.\ Rozowsky and B.\ Yan, \pl{401}{273}{97}.

\item \label{magnea}
K.\ Roland, \pl{289}{148}{92}; 
P.\ Di Vecchia, L.\ Magnea, A.\ Lerda,
R.~Marotta and R.\ Russo, \np{469}{235}{96}, 
\pl{388}{65}{96}.

\item \label{fqa}
M.G.\ Schmidt and C.\ Schubert, \pr{53}{2150}{96};
M.\ Reuter, M.G.~Schmidt and C.\ Schubert, Ann. Phys. 259 (1997) 313.

\item \label{reca}
C.\ Kim and V.P.\ Nair, \pr{55}{3851}{97}.

\item \label{aude}
A.\ Gehrmann-De Ridder and E.W.N.\ Glover, Durham preprint DTP/97/26 
(hep-ph/9707224).

\item \label{KST}
Z.\ Kunszt, A.\ Signer and Z. Tr\'ocs\'anyi, \np{420}{550}{94}.

\item \label{dsoft}
F.A.\ Berends and W.T.\ Giele, \np{313}{595}{89};
S.\ Catani, in Proceedings of the Workshop on {\it New Techniques for
Calculating Higher Order QCD Corrections}, report ETH-TH/93-01, Zurich (1992). 

\item \label{glover}
J.M.\ Campbell and E.W.N. Glover, Durham preprint DTP/97/82
(hep-ph/9710255).

\item \label{1loopcol}
Z.\ Bern, G.\ Chalmers, L.\ Dixon and D.A.\ Kosower, \prl{72}{2134}{94};
Z.\ Bern, L.\ Dixon, D.C.\ Dunbar and D.A.\ Kosower, \np{425}{217}{94};
Z.\ Bern, L.\ Dixon and D.A.\ Kosower, \np{437}{259}{95};
Z.\ Bern and G.~Chalmers, \np{447}{465}{95}.

\item \label{CC}
S.\ Catani and M.\ Ciafaloni, \np{249}{301}{85};
S.\ Catani, M.\ Ciafaloni and G.\ Marchesini, \np{264}{588}{86}.

\item \label{tHV}
G. 't~Hooft and M.\ Veltman, \np{44}{189}{72}.

\item \label{bollini}
G.\ Bollini and J.J.\ Giambiagi, Nuovo\ Cimento\ 12B (1972) 20;
J.F.\ Ashmore, Nuovo\ Cimento\ Lett. 4 (1972) 289;
G.M.\ Cicuta and E.\ Montaldi, Nuovo\ Cimento\ Lett. 4 (1972) 329.

\item \label{cdrir}
R.\ Gastmans and R.\ Meuldermans, \np{63}{277}{73}.

\item \label{Sie79}
W. Siegel, \pl{84}{193}{79} and 94B (1980) 37;
D.M.\ Capper, D.R.T.~Jones and P.\ van Nieuwenhuizen, \np{167}{479}{80};
L.V.\ Avdeev and A.A.~Vladi\-mirov, \np{219}{262}{83}.

\item \label{4dhs}
Z.\ Bern and D.A.\ Kosower, \np{379}{451}{92}.

\item \label{uni}
S.\ Catani, M.H.\ Seymour and Z.\ Tr\'ocs\'anyi, \pr{55}{6819}{97}.    

\item \label{KST2to2}
Z.\ Kunszt, A.\ Signer and Z.\ Tr\'ocs\'anyi, \np{411}{397}{94}.

\item \label{Kfactor}
J.\ Kodaira  and L.\ Trentadue, \pl{112}{66}{82}; 
G.P.\ Korchemsky and A.V.~Radyushkin, \np{283}{342}{87};
S.\ Catani, E.\ d'Emilio and  L.~Trenta\-due, \pl{211}{335}{88}; 
S.\ Catani and L.\ Trentadue, \np{327}{323}{89};
S.\ Catani, L.\ Trentadue, G.\ Turnock and B.R.\ Webber, \np{407}{3}{93}.

\item \label{CMW}
S.\ Catani, B.R.\ Webber and G.\ Marchesini, \np{349}{635}{91}.

\item \label{CFP}
G.\ Curci, W.\ Furmanski and R.\ Petronzio, \np{175}{27}{80}.

\item \label{ERT}
R.K. Ellis, D.A. Ross and A.E. Terrano, \np{178}{421}{81}.

\item \label{kramer}
K.\ Fabricius, I.\ Schmitt, G.\ Kramer and G.\ Schierholz, \zp{11}{315}{81}.

\end{enumerate}

\end{document}